\documentclass[aps,prd,notitlepage,nofootinbib,,nobibnotes,superscriptaddress,amsmath,amssymb,preprintnumbers]{revtex4-1}

\usepackage{graphicx} 
\usepackage{amsmath}
\usepackage{dcolumn}
\usepackage[colorlinks=true,linktocpage=true,%
  linkcolor=blue,citecolor=blue,urlcolor=blue]{hyperref}
\usepackage{color}
\usepackage{slashed}
\usepackage     [utf8]                  {inputenc}

\newcommand{\GeV}{\;\text{GeV}}
\newcommand{\TeV}{\;\text{TeV}}
\newcommand{\rmd}{\mathrm{d}}
\newcommand{\rme}{\mathrm{e}}
\newcommand{\rmi}{\mathrm{i}}

\newcommand{\Uf}{\tilde{U}}

\newcommand{\qbar}{\bar{q}}

\newcommand{\etag}{\eta_\gamma}

\newcommand{\LIR}{\Lambda_{\rm IR}}

\newcommand{\calN}{\mathcal{N}}

\newcommand{\xp}{\boldsymbol{x}_\perp} 

\newcommand{\khp}{\boldsymbol{k}_{1\perp}} 
\newcommand{\kgp}{\boldsymbol{k}_{\gamma\perp}} 

\newcommand{\kp}{\boldsymbol{k}_\perp}
\newcommand{\pp}{\boldsymbol{p}_\perp}
\newcommand{\qp}{\boldsymbol{q}_\perp}

\newcommand{\Pp}{\boldsymbol{P}_\perp} 

\begin{document}

\preprint{YITP-18-74}

\date{\today}

\title{Constraining unintegrated gluon distributions from inclusive photon production in proton-proton collisions at the LHC}
\author{Sanjin Beni\' c\footnote{On leave of absence from Department of Physics, Faculty of Science, University of Zagreb, Bijenička c. 32, 10000 Zagreb, Croatia}}
\affiliation{Yukawa Institute for Theoretical Physics,
Kyoto University, Kyoto 606-8502, Japan}
\author{Kenji Fukushima}
\affiliation{Department of Physics, The University of Tokyo,
                 7-3-1 Hongo, Bunkyo-ku, Tokyo 113-0033, Japan}
\author{Oscar Garcia-Montero}
\affiliation{Institut f\"{u}r Theoretische Physik,
                     Universit\"{a}t Heidelberg, Philosophenweg 16,
                     69120 Heidelberg, Germany}
\author{Raju Venugopalan}
\affiliation{Physics Department, Brookhaven National Laboratory,
              Bldg.\ 510A, Upton, NY 11973, USA}
              
\begin{abstract} 
We compute the leading order (LO) $qg\to q \gamma$ and next-to-leading order (NLO) $gg\to q{\bar q} \gamma$ contributions to inclusive photon production in proton-proton (p+p) collisions at the LHC. These channels provide the dominant contribution at LO and NLO for photon transverse momenta $k_{\gamma\perp}$ corresponding to momentum fractions of $x\leq 0.01$ in the colliding protons. Our computations, performed in the dilute-dense framework of the Color Glass Condensate effective field theory  (CGC EFT), show that the NLO contribution dominates at small-$x$ because it is sensitive to $k_\perp$-dependent unintegrated gluon distributions in both of the protons. 
We predict a maximal $10\%$ modification of the cross section at low $k_{\gamma\perp}$ as a direct consequence of the violation of $k_\perp$-factorization. 
The coherence effects responsible for this modification are enhanced in nuclei and can be identified from inclusive photon measurements in proton-nucleus collisions. We provide numerical results for the isolated inclusive photon cross section for  $k_{\gamma\perp}\leq 20$ GeV  in p+p collisions that can be tested in the future at the LHC.

\end{abstract}

\maketitle


Photon production in high energy hadron-hadron collisions provides an excellent tool to probe the small-$x$ structure of hadron wavefunctions which are dominated by Fock states containing a large abundance of gluons. Their dynamics is described by the Color Glass Condensate (CGC) effective field theory (EFT)~\cite{Iancu:2003xm,Gelis:2010nm}. The dominant contribution to inclusive photon production at small-$x$, within the dilute-dense framework of CGC, is from the $qg\to q\gamma$ channel; it has been computed in several papers~\cite{Kopeliovich:1998nw,Gelis:2002ki,Baier:2004tj,Kopeliovich:2007fv}, with further applications to proton-proton (p+p) \cite{Kopeliovich:2007yva,Kopeliovich:2009yw,Rezaeian:2009it} and proton-nucleus (p+A)  collisions~\cite{JalilianMarian:2005gm,JalilianMarian:2012bd,Rezaeian:2012wa,Basso:2012nb,Rezaeian:2016szi,Ducloue:2017kkq}. Since the occupancy of gluons in the target is of order 
$1/\alpha_S$ at small-$x$, where $\alpha_S$ is the QCD coupling, the quark from the proton scatters coherently off a gluon shockwave in the target. This channel is dominant in the  fragmentation region where the hard photon is emitted off a large-$x$ valence quark scattering off the small-$x$ gluons in the target.

In \cite{Benic:2016uku}, we computed the next-to-leading order (NLO) channel $gg\to q{\bar q} \gamma$ channel in the CGC EFT. For photon rapidities that are close to the central rapidity region of the collision, this process dominates over other contributions at this order~\cite{Benic:2016yqt,Altinoluk:2018uax,Altinoluk:2018byz}. It can be visualized as a fluctuation of a gluon from one of the protons into a quark-antiquark pair that scatters off the gluon shock wave of the other proton. Alternately, this gluon can first scatter off the shockwave before fluctuating into the quark-antiquark pair. In either case, the pair can emit a hard photon. If one probes small-$x$ values in either proton of $x\leq 0.01$, this NLO process will dominate over the stated LO contribution because the large gluon density in the proton overcompensates for the $\alpha_S$ suppression in the NLO cross-section arising from the splitting of the gluon into the quark-antiquark pair. 

Our computation was performed within the dilute-dense approximation in the CGC EFT~\cite{Blaizot:2004wu,Blaizot:2004wv}, wherein one computes pair production (and subsequent photon emission) by solving the Dirac equation in the classical background field generated in the scattering process to lowest order in $\rho_p/k_{p\perp}^2$ and to all orders in $\rho_t/k_{t\perp}^2$. Here $\rho_p$ ($\rho_t$) are the color charge densities in the projectile (target) proton, and $k_{p\perp}$ ($k_{t\perp}$) are the associated transverse momenta. This approximation is strictly valid in the forward rapidity region where the momentum fraction $x_t$ of the parton from the ``target"  proton is much smaller than $x_p$, the momentum fraction of the parton from the ``projectile" proton. Note that for these assumptions to be a priori robust, even the projectile parton should have $x_p\leq 0.01$. In our computations, we will cover kinematic regimes that will fall outside this preferred kinematic regime; the systematic uncertainties of the computation increase in that case due to the increased contributions of other channels and/or higher order effects.  We note that the computation of heavy quark pairs $gg\to q{\bar q} $ in this framework (which, by Low's theorem, is a limit of our results in the limit of $k_{\gamma\perp}\rightarrow 0$) has been applied, with considerable success, to describe heavy quarkonium production in p+p collisions at RHIC and the LHC~\cite{Ma:2014mri}, in  p+A collisions at both colliders~\cite{Ma:2015sia,Qiu:2013qka,Ma:2017rsu} and more recently, high multiplicity p+p and p+A collisions~\cite{Ma:2018bax}. In the latter case, the framework employed here also gives very good agreement with multiplicity distributions at the LHC~\cite{Dumitru:2018gjm}. 

In this work, we will extend the application of the dilute-dense CGC EFT to single inclusive photon production in p+p collisions at the LHC energies. The photon data available thus far is from ATLAS and CMS \cite{Chatrchyan:2012vq,Khachatryan:2010fm,Chatrchyan:2011ue,Aad:2010sp,Adriani:2017jys} where $k_{\gamma\perp} >  20 $ GeV, with the exception of one data point extending below $20$ GeV. While these values of the photon $k_{\gamma\perp}$ are too hard to be directly sensitive to small $x$ dynamics in the proton wavefunction, it is anticipated that ALICE will measure lower-$k_{\gamma\perp}$ photons. Especially promising are the forward LHC upgrades \cite{N.Cartiglia:2015gve}, such as the LHCf \cite{Adriani:2006jd} and the proposed ALICE FoCal \cite{Peitzmann:2016gkt} upgrades. 

As a reasonable estimate of the kinematic reach of the CGC EFT, we will impose the condition that the average $x$ in the target proton is $x < 0.01$;  for LHC energies, this corresponds approximately to $k_{\gamma\perp} \lesssim 20$ GeV at mid-rapidities. The CGC-based formulas, as explicitly laid out in the following, have a systematic $k_\perp$-factorized (or dilute-dilute) limit, wherein the cross--section is factorized into the product of unintegrated gluon distributions (UGDs) in each of the protons. Deviations from this $k_\perp$ limit increase with increasing values of 
either $\rho_p/k_\perp^2$ or $\rho_t/k_\perp^2$, with maximal contributions coming from $k_\perp\sim Q_s$, where $Q_s$ is the saturation scale in the projectile or target at a given $x$. 
Thus in the CGC framework one can extract information on the UGD distributions by comparing the computed inclusive photon distributions to data as well as quantify saturation effects by looking for systematic deviations from the $k_\perp$ factorized formalism along the lines predicted in the CGC EFT.

We begin by summarizing the CGC formulas for the LO and the NLO processes to explain our notations, approximations, and details of the  numerical computation. 
The cross-section\footnote{We use the following abbreviations;
  $\int_{\qp} \equiv \int \frac{\rmd^2 \qp}{(2\pi)^2}$ and
  $\int_{\xp} \equiv \int \rmd^2 \xp$.} in the dilute-dense approximation
of the LO process $qg\to q(q)\gamma(k_\gamma)$ in the dilute-dense collision is
given by~\cite{Kopeliovich:1998nw,Gelis:2002ki,Baier:2004tj,Kopeliovich:2007fv}
\begin{equation}
\begin{split}
&\frac{\rmd\sigma^{\rm LO}}{\rmd^2 \kgp \rmd \etag} = S_\perp \sum_f\frac{\alpha_e q_f^2}{16\pi^2}\int_{\qp}\int_{x_{p,{\rm min}}}^1 \rmd x_p\, f^{\rm val}_{q,f}(x_p,Q^2)\,\tilde{\calN}_{t,Y_t}(\qp + \kgp)\\
&\times\frac{1}{q^+ l^+}\Bigg\{-4m_f^2\left[\frac{l^{+2}}{(q\cdot k_\gamma)^2}+\frac{q^{+2}}{(l\cdot k_\gamma)^2} + \frac{k_\gamma^{+2}}{(l\cdot k_\gamma)(q\cdot k_\gamma)}\right]+4\left(l^{+2} + q^{+2}\right)\left[\frac{l\cdot q}{(l\cdot k_\gamma)(q\cdot k_\gamma)}+\frac{1}{q\cdot k_\gamma}-\frac{1}{l\cdot k_\gamma}\right]\Bigg\}\,,
\end{split}
\label{eq:inclo}
\end{equation}
where $f^{\rm val}_{q,f}(x_p,Q^2)$ is the valence quark distribution
function with $Q^2 = {\rm max}(\qp^2,\kgp^2)$, $S_\perp$ is the
transverse proton size, and $m_f$ is the quark mass for flavor $f$.
The gluon shockwave in the dense target is represented by the dipole forward scattering amplitude,
\begin{equation}
  \tilde{\calN}_{t,Y_t}(\kp) = \frac{1}{N_c}\int_{\xp}
  \rme^{\rmi\kp\cdot\xp}\mathrm{tr}_c\langle\Uf(\xp)\Uf^\dag(0)\rangle_{Y_t}\,.
\label{eq:dip}
\end{equation}

In the above, the rapidity of the dense target is $Y_t = \log(1/x_t)$
with $x_t = \sqrt{2/s} \, (q^- + k_\gamma^-)$
and $\Uf(\xp)$ is a fundamental lightlike Wilson line.
The light cone momenta of the incoming
quark are $l^+ = \sqrt{\frac{s}{2}}x_p$ and $l^- = m_f^2/(2l^+)$,
those of the final state quark are: $q^+ = l^+ - k_\gamma^+$ and
$q^- = (\qp^2 + m_f^2)/(2q^+)$. Finally, those of the photon are 
$k_\gamma^{\pm} = k_{\gamma\perp} \rme^{\pm\etag}/\sqrt{2}$.
We note that $q^+ > 0$ leads to $x_p \ge x_{p,{\rm min}}$ with
$x_{p,{\rm min}} = \sqrt{2} k_\gamma^+/\sqrt{s}$.

For inclusive photon production at NLO in $\alpha_S$, as noted, there are three different channels in the gluon shockwave background of the target proton:
$qg \to qg\gamma$ \cite{Altinoluk:2018uax,Altinoluk:2018byz},
$gg \to q^\ast \qbar^\ast \to \gamma$ \cite{Benic:2016yqt}, and
$gg \to q\qbar\gamma$ \cite{Benic:2016uku}.
The collinearly enhanced contributions in the tree-level process $qg\to qg\gamma$ are
contained in the LO with evolved valence quark distributions, 
while the $gg\to q^\ast\qbar^\ast \to \gamma$ channel is suppressed by the virtual
$q^\ast \qbar^\ast$ phase space and flavor
cancellation~\cite{Benic:2016yqt}.  In the present work,  we will consider
the region close to mid-rapidity of $0 <  Y_p < 2.5$ where the tree-level $gg\to q\qbar \gamma$ channel is the
dominant contribution. The $qg\to qg \gamma$ channel, which may be expected to play an important role in the very forward region of the dilute projectile,
will not be discussed in the following.

The inclusive cross section of the photon production from the 
$gg\to q (q) + \qbar (p) + \gamma (k_\gamma)$ channel can be expressed as~\cite{Benic:2016uku},
\begin{equation}
\begin{split}
  \frac{\rmd \sigma^{\rm NLO} }{\rmd^2 \kgp \rmd \etag} &=
  S_\perp \sum_f \frac{\alpha_e \alpha_S N_c^2 q_f^2}{64\pi^4 (N_c^2 - 1)} \int_{\eta_q \eta_p}\int_{\qp \pp \khp \kp}\frac{\varphi_p(Y_p,\khp)}{\khp^2}\tilde{\calN}_{t,Y_t}(\kp)\tilde{\calN}_{t,Y_t}(\Pp - \khp-\kp)\\
&\quad\times \bigl[ 2\tau_{g,g}(\khp;\khp)
  +4\tau_{g,q\qbar}(\khp;\kp,\khp) +2\tau_{q\qbar ,q\qbar}(\kp,\khp;\kp,\khp)\bigr]\,,
\end{split}
\label{eq:incnlo}
\end{equation}
where $\Pp=\qp+\pp+\kgp$ and the rapidities are
$Y_{p,t}=\log(1/x_{p,t})$ with 
\begin{equation}
x_p=\sqrt{\frac{2}{s}} \, (q^++p^++k_\gamma^+) \quad \mathrm{and} \quad  
x_t=\sqrt{\frac{2}{s}} \, (q^-+p^-+k_\gamma^-)\,
\label{eq:xs}
\end{equation}
Here the light-cone momenta of an on-shell particle with 4-momentum $p$ are given by 
\begin{equation}
p^{\pm} = \frac{1}{\sqrt{2}}\sqrt{\pp^2 + m^2}\,\, \exp(\pm \eta_p)\,.
\label{eq:lcm}
\end{equation}

The unintegrated gluon distribution (UGD) in the
dilute projectile $\varphi_p(Y_p,\khp)$ is defined as
\begin{equation}
\varphi_p(Y_p,\khp) \equiv S_\perp \; \frac{N_c \,\khp^2}{4\alpha_S}\calN_{p,Y_p}(\khp) \,,
\label{eq:dipa}
\end{equation}
where $\calN_{p,Y_p}(\kp)$, the dipole amplitude  is expressed in terms of the
adjoint lightlike Wilson line $U(\xp)$ as 
\begin{equation}
\calN_{p,Y_p}(\kp) = \frac{1}{N_c}\int_{\xp}\rme^{\rmi\kp\cdot\xp}\mathrm{tr}_c\langle U(\xp)U^\dag(0)\rangle_{Y_p}\,.
\end{equation}
The product of fundamental dipoles in Eq.~\eqref{eq:incnlo}, to $O(1/N_c^2)$ in a large-$N_c$ expansion, represents general multigluon correlators describing the dense target; these too can be represented formally as UGDs~\cite{Blaizot:2004wv}.

The square brackets in Eq.~\eqref{eq:incnlo} contain the hard factors
for this process, where $\tau_{n,m}$ with $n,m\in\{g,q\qbar\}$ represents the Dirac trace,
\begin{equation}
\tau_{n,m} \equiv {\rm tr}\bigl[(\slashed{q} + m_f)T^\mu_n (m_f - \slashed{p})\gamma^0 T'^\dag_{m,\mu}\gamma^0\bigr]\,,
\label{eq:tnm}
\end{equation} 
with Dirac matrix products $T^\mu_n$ as specified in~\cite{Benic:2016uku}. 

If $k_{\gamma\perp}$ is much larger than the typical momenta
exchanged from the dense target, namely $k_\perp$ and
$|\Pp - \khp - \kp|$, Eq.~\eqref{eq:incnlo} simplifies to a
$k_\perp$-factorized expression,
\begin{equation}
\begin{split}
\frac{\rmd \sigma^{\rm NLO}_{k_\perp \text{-fact}}}{\rmd^2\kgp \rmd\etag}
&= S_\perp \sum_f \frac{\alpha_e \alpha_S N_c^2 q_f^2}{64\pi^4 (N_c^2 - 1)} \int_{\eta_q \eta_p}\int_{\qp \pp \khp}\frac{\varphi_p(Y_p,\khp)}{\khp^2}\,\calN_{t,Y_t}(\Pp - \khp)\\
&\quad\times \bigl[2\tau_{g,g}(\khp)
  +\tau_{q,q}(\khp) +\tau_{\qbar,\qbar}(\khp) + 2\tau_{g,q}(\khp) + 2\tau_{g,\qbar}(\khp)\bigr]\,,
\end{split}
\label{eq:incnlokt}
\end{equation}
where $\tau_{n,m}$ takes the same form as in Eq.~\eqref{eq:tnm} for
$n,m\in\{g,q,\qbar\}$ with the additional Dirac structures $T^\mu_q$ and $T^\mu_{\qbar}$ also specified as in \cite{Benic:2016uku}.
In this limit, the higher twist contributions in the projectile and the target gluon distributions are small corrections and the $k_\perp$-factorized formula \eqref{eq:incnlokt} smoothly turns into the leading twist, or dilute-dilute, approximation of Eq.~\eqref{eq:incnlo}.

It is crucial to note that we employ only the valence quark distribution in
Eq.~\eqref{eq:inclo} and not the sea quark distribution. The reason for this is as follows. When valence quarks radiate gluons, the collinear gluon emissions are enhanced and generate a gluon distribution. If these collinear gluons subsequently radiate sea quarks, and the photon is emitted off a sea quark leg, where the incoming sea quark is collinear to the gluon, that contribution, after integration over the phase space of the spectators, will give a contribution that formally will have the structure of our LO result. However, this result is entirely contained in our NLO expression and can be obtained by taking the appropriate collinear limits thereof. Hence including sea quarks in the LO computation would amount to double counting their contribution. We therefore perform the flavor summation in Eq.~\eqref{eq:inclo} only over the valence $u$ and $d$ quarks, while the flavor summation in Eq.~\eqref{eq:incnlo} and \eqref{eq:incnlokt} runs over $u$, $d$, $s$, $c$ and $b$ quarks.

Prompt photon production includes both the direct photon component described by the above formulae as well as the contribution from fragmentation photons that we do not compute here. 
Experimentally, the two contributions can be separated by imposing an isolation cut along lines similar to that proposed in \cite{Frixione:1998jh}; while this minimizes the fragmentation contribution, it does not eliminate it completely and this uncertainty is part of the quoted experimental systematic errors. 
We will adopt here the same isolation cut as used in the experiments to compare our
results to the data.  The above formulas must be convoluted with
\begin{equation}
 \theta\Bigl(\sqrt{(\etag - \eta)^2 + (\phi_\gamma- \phi)^2}
 - R\Bigr)\,,
 \label{eq:isocut}
\end{equation}
where $\theta(x)$ is the step function, $\eta$, $\phi$ are respectively the 
rapidity and the azimuthal angle of either\footnote{Hence, for the $gg\to q\qbar\gamma$ channel one needs to 
insert two step functions.} $q$ or $\qbar$,  while $\etag$ and
$\phi_\gamma$ denote the rapidity and the azimuthal angle of the
photon.  The CMS and the ATLAS experiments use $R = 0.4$, estimating
the remaining fragmentation component to $10\%$ of the total cross
section~\cite{Ichou:2010wc,dEnterria:2012kvo}.  We use $R=0.4$ throughout this paper.

We will now present some of the numerical details in our computation of Eqs.~\eqref{eq:inclo}, \eqref{eq:incnlo} and \eqref{eq:incnlokt}.
For the valence quark distribution, we use the CTEQ6M
set~\cite{Pumplin:2002vw}.
The small-$x$ evolution of the dipole distributions is obtained from the running coupling Balitsky-Kovchegov (rcBK) \cite{Balitsky:1995ub,Kovchegov:1999yj}, which is 
a good approximation to the general expression for the dipole forward scattering amplitude given by the Balitsky-JIMWLK hierarchy  \cite{Balitsky:1995ub,JalilianMarian:1997jx,JalilianMarian:1997dw,Iancu:2000hn,Iancu:2001ad}.
In solving the rcBK equation numerically, the initial condition for the dipole amplitude 
at $x_0=0.01$ is given by the McLerran-Venugopalan (MV) model with anomalous dimension $\gamma=1$, 
the saturation momentum at the initial $x_0$ of $Q^2_0=0.2\GeV^2$, and the IR cutoff for
the running couping $\LIR=0.241\GeV$--see \cite{Dusling:2009ni} for details of the rcBK initial conditions. 
With the initial condition fixed, the rcBK equation  is solved to determine the dipole amplitude for $x< x_0$. 
For  $x > x_0$, we use the extrapolation suggested in Ref.~\cite{Ma:2014mri} wherein the UGD can be matched to the CTEQ6M gluon distribution. The matching procedure 
fixes the proton radius $R_p$, to $R_p = 0.48$ fm, or equivalently $S_\perp = \pi\,R_p^2 = 7.24$ mb.  Note that this value of $R_p$ is quite close to that extracted from saturation model fits to exclusive DIS data~\cite{Rezaeian:2012ji}. In our computations, we will take quark masses to be typically 
$m_u = m_d = 0.005\GeV$, $m_s = 0.095\GeV$, $m_c = 1.3\GeV$ and $m_b = 4.5\GeV$. We will discuss later the effects of varying the parameters on model to data comparisons. 
 

\begin{figure}
  \centering
  \includegraphics[width=0.5\textwidth]{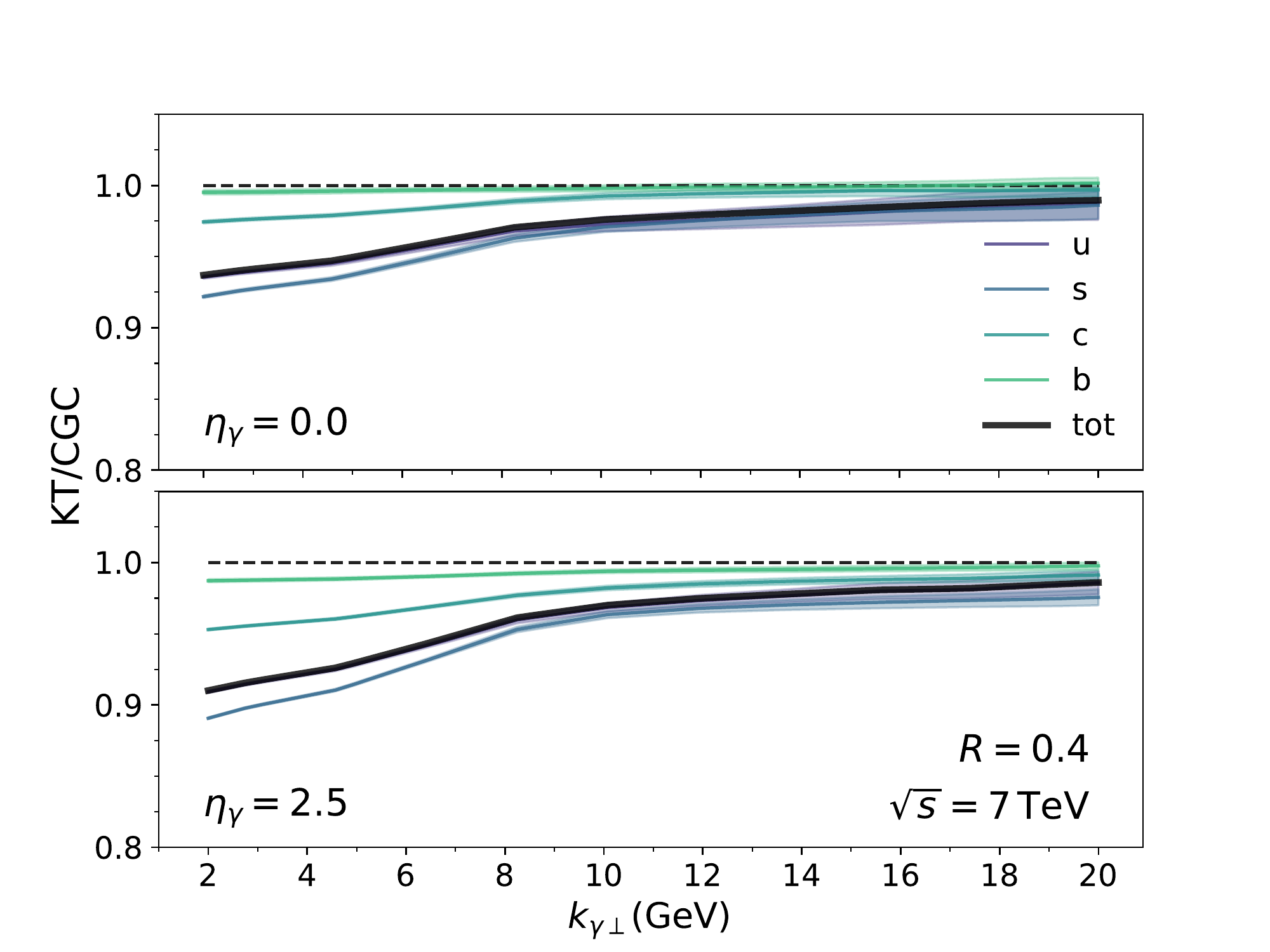}
  \caption{Ratios of the $k_\perp$-factorized results to the full CGC
    results as a function of $k_{\gamma\perp}$ at
    $\sqrt{s}=7\;\text{TeV}$ with the isolation cut $R=0.4$.  The
    upper panel is for the photon rapidity $\eta_\gamma=0$ and the
    lower for $\eta_\gamma=2.5$. The band represents the error estimate from performing multidimensional integrals using the  VEGAS Monte Carlo integration routine.}
\label{fig:cgcvskt}
\end{figure}

Evaluating the full CGC formula for the single inclusive photon cross-section as a function of photon transverse momenta $k_{\gamma\perp}$ and rapidity $\eta_\gamma$ in Eq.~\eqref{eq:incnlo} involves performing 10-dimensional 
integrations while the simpler $k_\perp$-factorized approximation in Eq.~\eqref{eq:incnlokt} involves 8-dimensional integrations. Such multidimensional integrations are most efficiently performed by employing the VEGAS Monte Carlo (MC) algorithm. For the $k_\perp$-factorized integral, $10^8$ points were used to sample the approximate distribution of the integrand, 
until convergence with a significance of $\chi=0.3$ was obtained. For the CGC calculations, we used the same algorithm but sampled the integrand with $10^9$ points. As a numerical check of our computation, we confirmed that in the small $k_{\gamma\perp}$ limit the NLO result reproduces the soft photon theorem--see
Eqs.~(B.7)-(B.11) in Ref.~\cite{Benic:2016uku}.

At low to moderate $k_{\gamma\perp}$, the full-CGC computation of the inclusive photon cross section based on \eqref{eq:incnlo} breaks $k_\perp$-factorization. This is also the case for  inclusive quark production, as shown previously~\cite{Fujii:2005vj}.
Our results for $k_\perp$-factorization breaking are shown in Fig.~\ref{fig:cgcvskt}, where we plot the ratio of the
full CGC inclusive photon cross-section to the \ $k_\perp$-factorized  cross-section 
at $\sqrt{s}=7 \TeV$ and $R=0.4$. The results are plotted  for central and forward photon
rapidities, for individual flavor contributions, and for the net sum over flavors.
The breaking of $k_\perp$-factorization is greater for forward rapidities and for decreasing quark mass, with negligible breaking of $k_\perp$-factorization observed for the heaviest flavor. Quantitatively, the breaking is maximally $\sim 10\%$ breaking at the lowest $k_{\gamma\perp}$, approaching  unity for
$k_{\gamma\perp} \gtrsim 20\GeV$. As suggested by the discussion in \cite{Fujii:2006ab}, when $k_{\gamma\perp}$ is small, the quark-antiquark pair are more likely to both scatter off the gluon shockwave in the target; the $k_\perp$-factorized configuration, where multiple scattering of both the quark and antiquark does not occur, is therefore suppressed. As also suggested by Fig.~\ref{fig:cgcvskt}, the reverse is true at large $k_{\gamma\perp}$.


\begin{figure}
  \centering
  \includegraphics[width=0.45\textwidth]{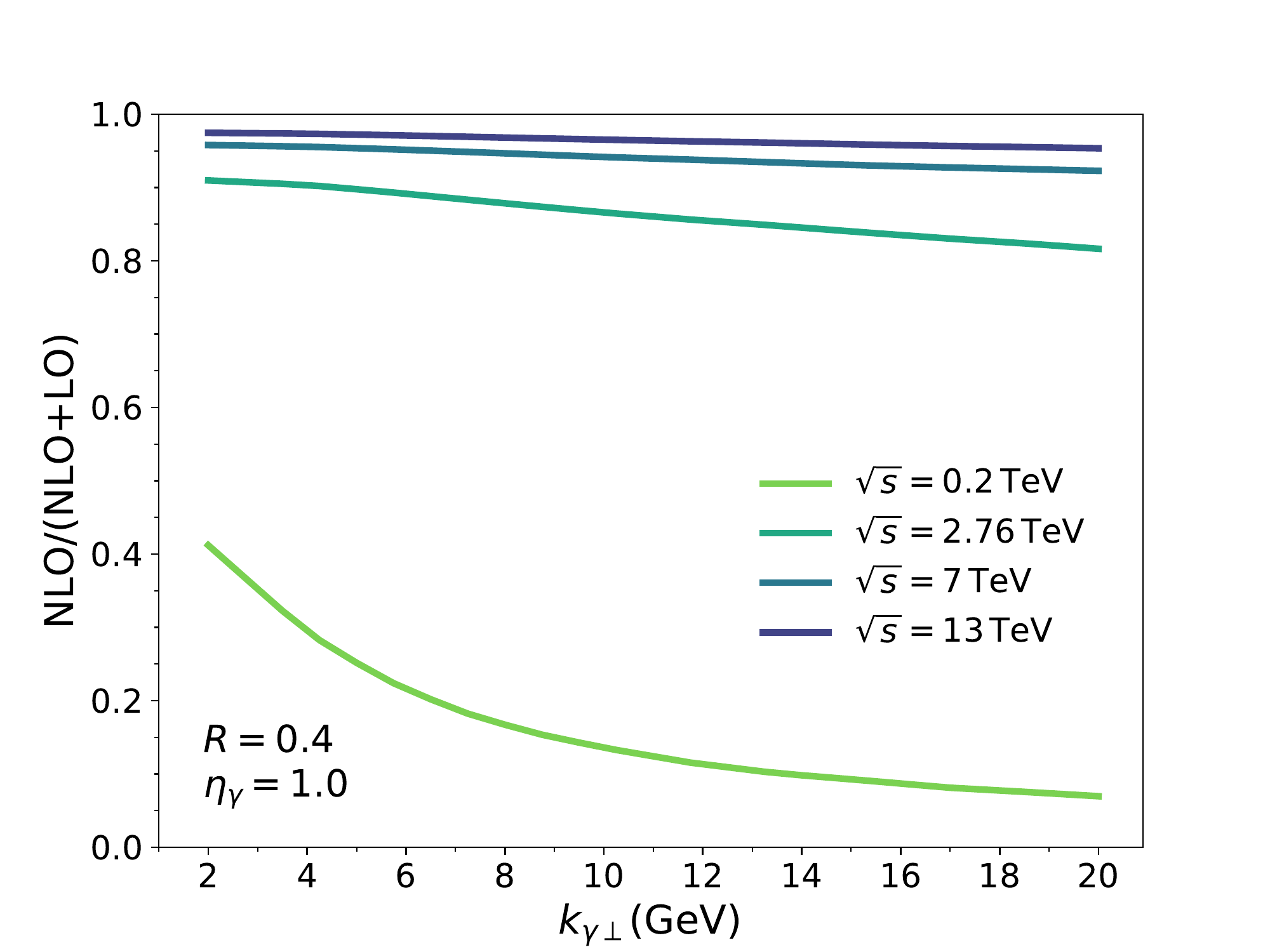}
   \includegraphics[width=0.45\textwidth]{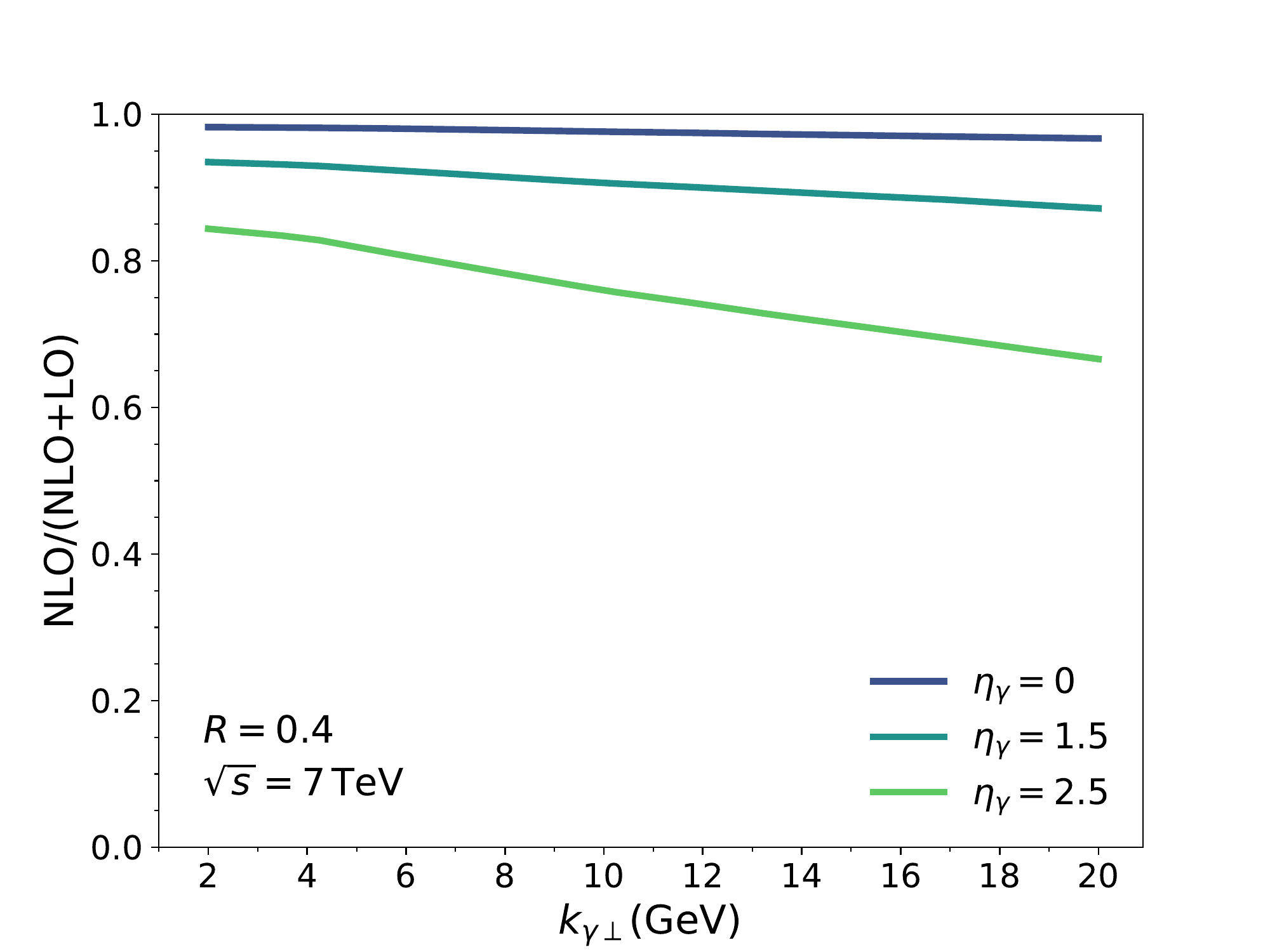}
  \caption{Fraction of the inclusive photon cross section from the NLO
    $gg\to q\qbar\gamma$ channel relative to the total NLO+LO contribution, as a
    function of $k_{\gamma\perp}$. Here, and in subsequent plots, the NLO computation was performed employing the $k_\perp$-factorized formula Eq.~\eqref{eq:incnlokt}. The left panel shows the collision
    energy dependence at $\sqrt{s} = 0.2, 2.76, 7, 13\TeV$ for
    $\eta_{\gamma} = 1.0$.  The right panel shows the photon
    rapidity dependence at $\eta_{\gamma} = 0, 1.5, 2.5$ for
    $\sqrt{s}=7\TeV$. In both cases, $R = 0.4$.}
\label{fig:frac}
\end{figure}

Next, to illustrate the importance of the NLO
($gg\to q\qbar \gamma$) channel quantitatively relative to the LO ($qg\to q\gamma$)
channel, we plot in Fig.~\ref{fig:frac} the NLO / (NLO+LO) fraction as a function of
$k_{\gamma\perp}$.  The left panel shows
the collision energy dependence of the ratio for $\sqrt{s}=0.2$, $2.76$, $7$ and $13\TeV$
with $\eta_\gamma = 1.0$.  We observe that the NLO fraction of the inclusive photon cross-section at the highest RHIC energy of $\sqrt{s} = 0.2\TeV$ is quite small, $\sim 10\%$. This is because, for the relevant $k_{\gamma\perp}$, quite large values of $x$ are probed in the proton where the gluon distribution does not dominate over that of valence quark distributions. However, already at $\sqrt{s} = 2.76\TeV$, the NLO contribution is more than $60\%$ even for the largest values of  $k_{\gamma\perp}$ shown, and increasing the center-of-mass energy to $\sqrt{s} = 7 \TeV$ and $13 \TeV$ enhances the NLO contribution to more than $\sim 90\%$. These results confirm that at LHC energies gluons dominate the proton wavefunction, even for photons with $k_{\gamma\perp}=20$ GeV. The right panel shows the ratio for photon rapidities of $\etag=0$, $1.5$, $2.5$ at a fixed $\sqrt{s}=7\TeV$. The NLO contribution dominates completely at central rapidities and supplies $50$\% of the cross-section even at $\etag = 2.5$ and $k_{\gamma\perp}=20$ GeV.


\begin{figure}
  \centering
  \includegraphics[width=0.45\textwidth]{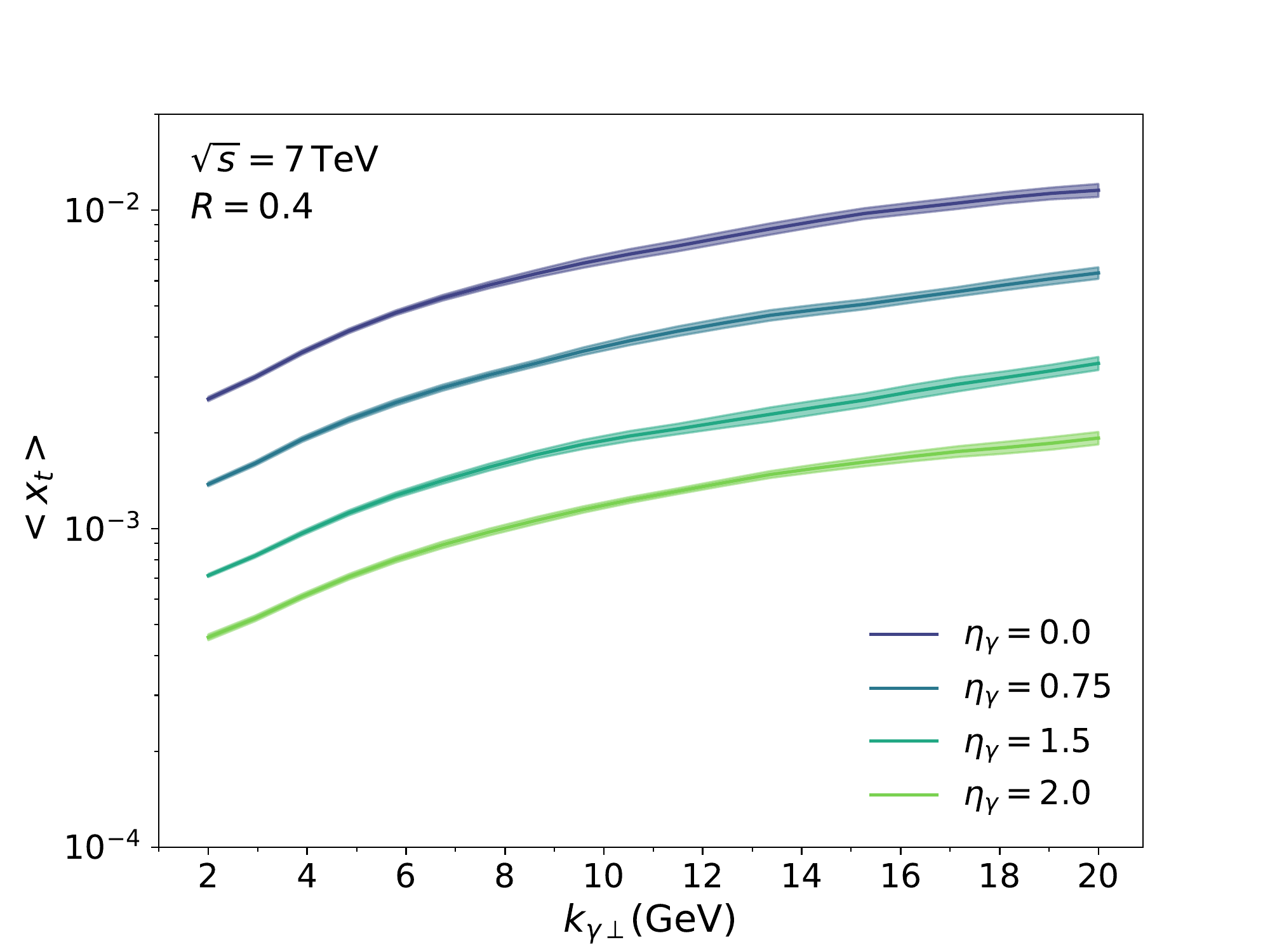}
   \includegraphics[width=0.45\textwidth]{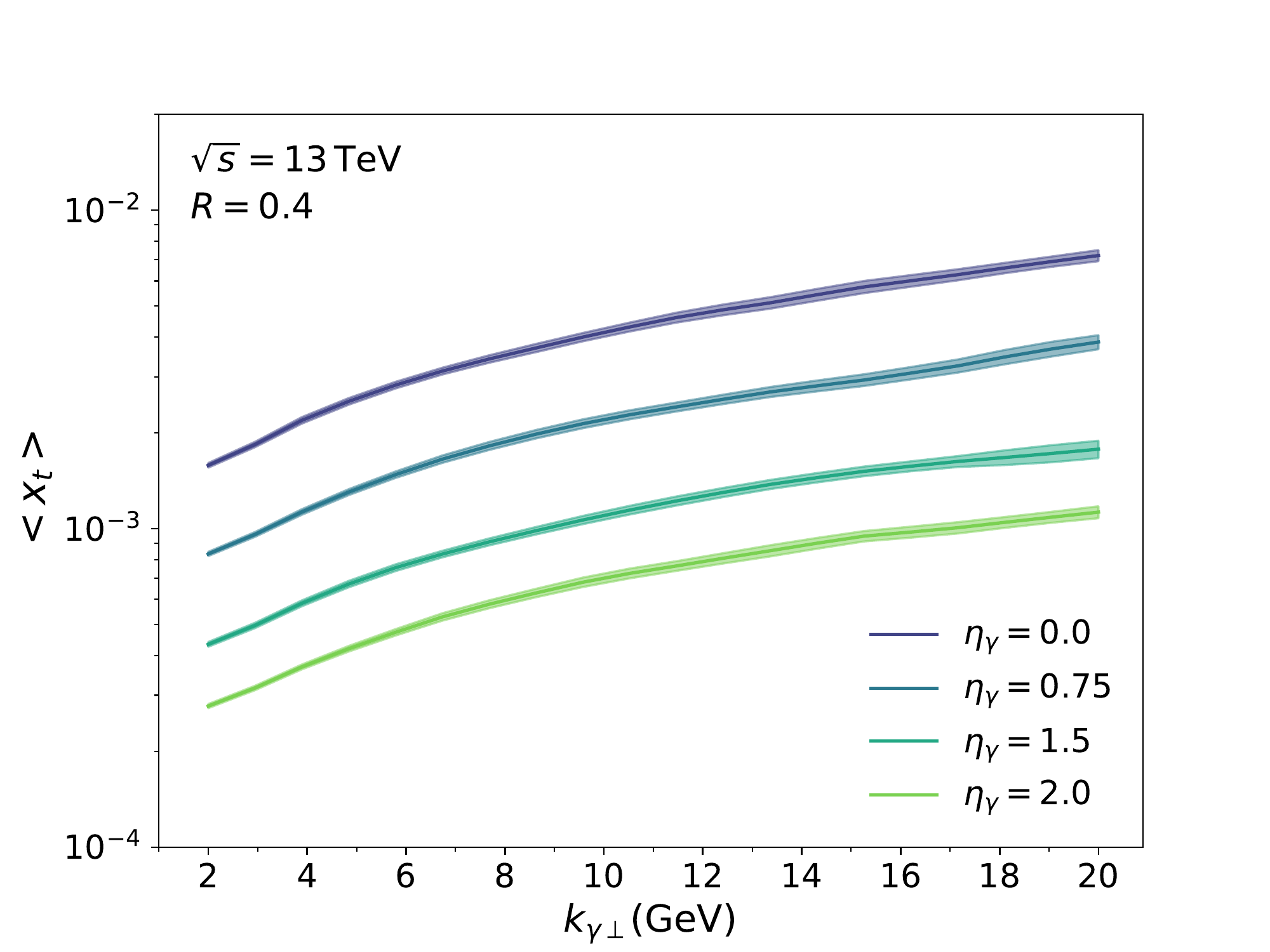}
  \caption{Left (right) panel shows $\langle x_t\rangle$,  the average value of $x_t$, in the target proton as a function of $k_{\gamma\perp}$ at $\sqrt{s} = 7 \TeV$ ($13 \TeV$). The different curves correspond to $\eta_\gamma=0.0$, $0.75$, $1.5$ and $2.0$. In both cases, $R = 0.4$.}
\label{fig:xav}
\end{figure}

A significant source of theoretical uncertainty in our computations are the contributions from the large $k_{\gamma\perp}$ region. Starting from $k_{\gamma\perp} \sim 10$ GeV, the small-$x$ logs compete with transverse momentum logs $\log(k_\perp^2/\Lambda_{\rm QCD}^2)$ associated with DGLAP evolution\footnote{According to a recent estimate \cite{Ball:2017otu}, small-$x$ effects in DIS become important for $\log 1/x \geq 1.2 \log Q^2/\Lambda_{\rm QCD}^2$. This estimate is process dependent and may be different in the case of inclusive photon production.} where a matching between the two formalisms becomes necessary. We will therefore show our results for $k_{\gamma\perp} \leq 20$ GeV where the average value of $x_t$ is $\langle x_t\rangle \leq 0.01$, as demonstrated on Fig.~\ref{fig:xav}.
For a systematic approach to this matching \cite{Catani:1990eg} it will be necessary to include higher order corrections to our framework. In addition to higher order contributions in QCD evolution and in the matrix elements, there are uncertainties in the extraction of the transverse area $S_\perp$. Though $S_\perp$ is constrained from the matching to parton distributions at large $x$, there can easily be $50$\% uncertainties in the overall cross-section that are absorbed by the extraction of the $K$-factor from comparison of the computed cross-sections to data. Until we can quantify the sources contributing to this $K$-factor separately, we should understand these sources of uncertainty as being ``bundled" together in the value extracted.

We should note further that there are other sources of uncertainty. We previously mentioned the $1/N_c^2$ corrections in using the BK truncation of the JIMWLK hierarchy. In practice, these are significantly smaller, specially so in the regime where $k_\perp$-factorization is applicable. Another source of systematic uncertainty are the values of the quark masses. Varying the quark masses in the ranges $m_{u,d} = 0.003-0.007$ GeV, $m_s = 0.095-0.15$ GeV, $m_c = 1.3-1.5$ GeV and $m_b = 4.2 - 4.5$ GeV, we observed that the cross section for $10 \, {\mathrm{GeV}} < k_{\gamma\perp} < 50 \, {\mathrm{GeV}}$ varies by $5-10\%$ for the light $u$, $d$, and $s$ quarks, while the heavier $c$ and $b$ quarks have small variations of order $0-5\%$. There is an overall degree of uncertainty in performing the Monte Carlo integrals, which is quantified by the error estimate of the VEGAS algorithm. This error estimate for the $k_\perp$-factorized inclusive cross-section is the range of  $0-5\%$ for all flavors. Based on these sources of uncertainty, we have included a systematic error band of 15\% in comparisons to data.

\begin{figure}
  \begin{center}
  \includegraphics[scale = 0.5]{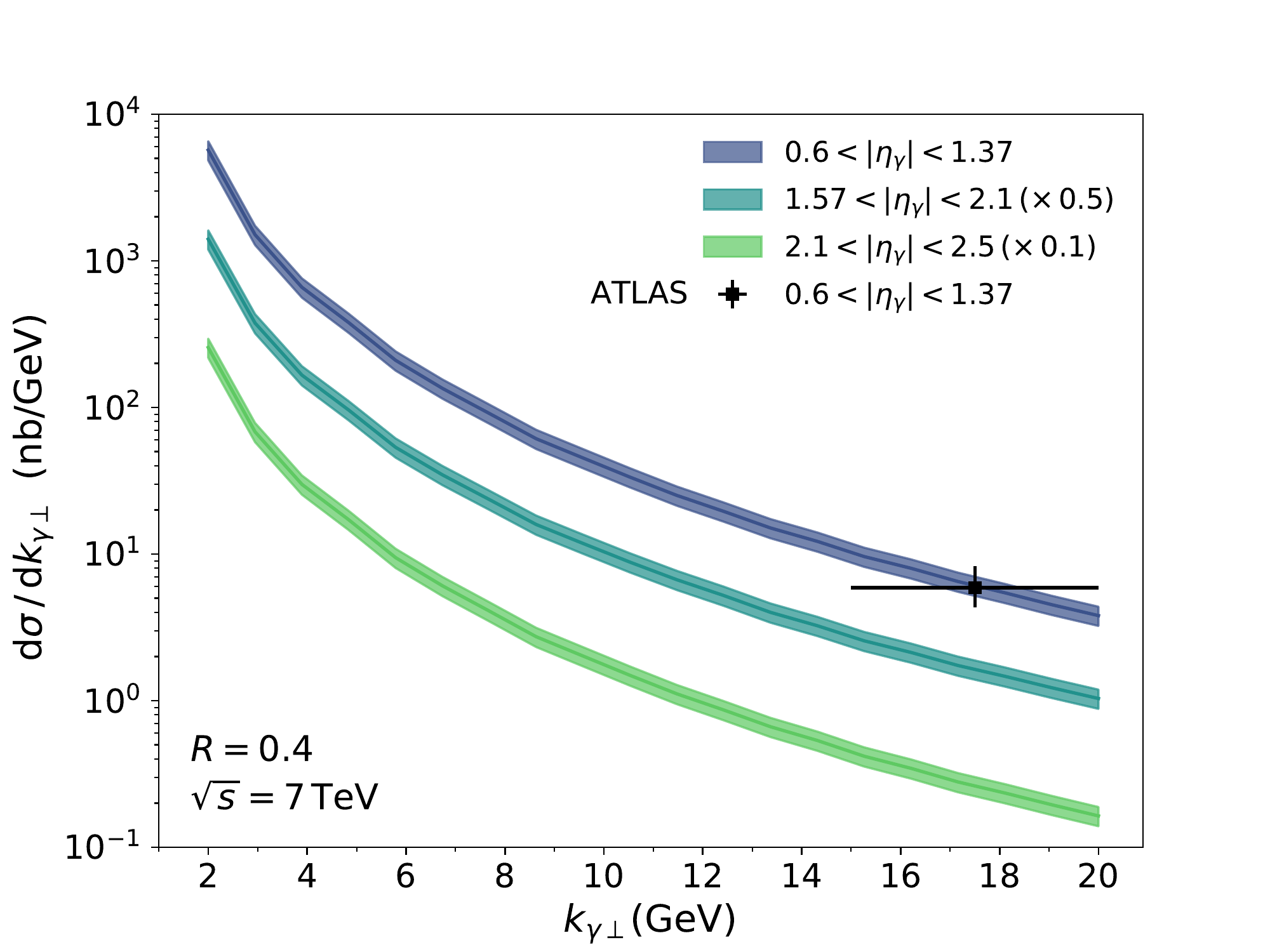}
  \end{center}
  \caption{Numerical results for the p+p photon data at $\sqrt{s}=7$ TeV across several rapidity bins. The central lines are obtained by multiplying our numerical results with a $K$-factor of $K = 2.4$. The data point is from the ATLAS experiment \cite{Aad:2010sp}.}
  \label{fig:atlascms7TeV}
\end{figure}

\begin{figure}
  \begin{center}
  \includegraphics[scale = 0.5]{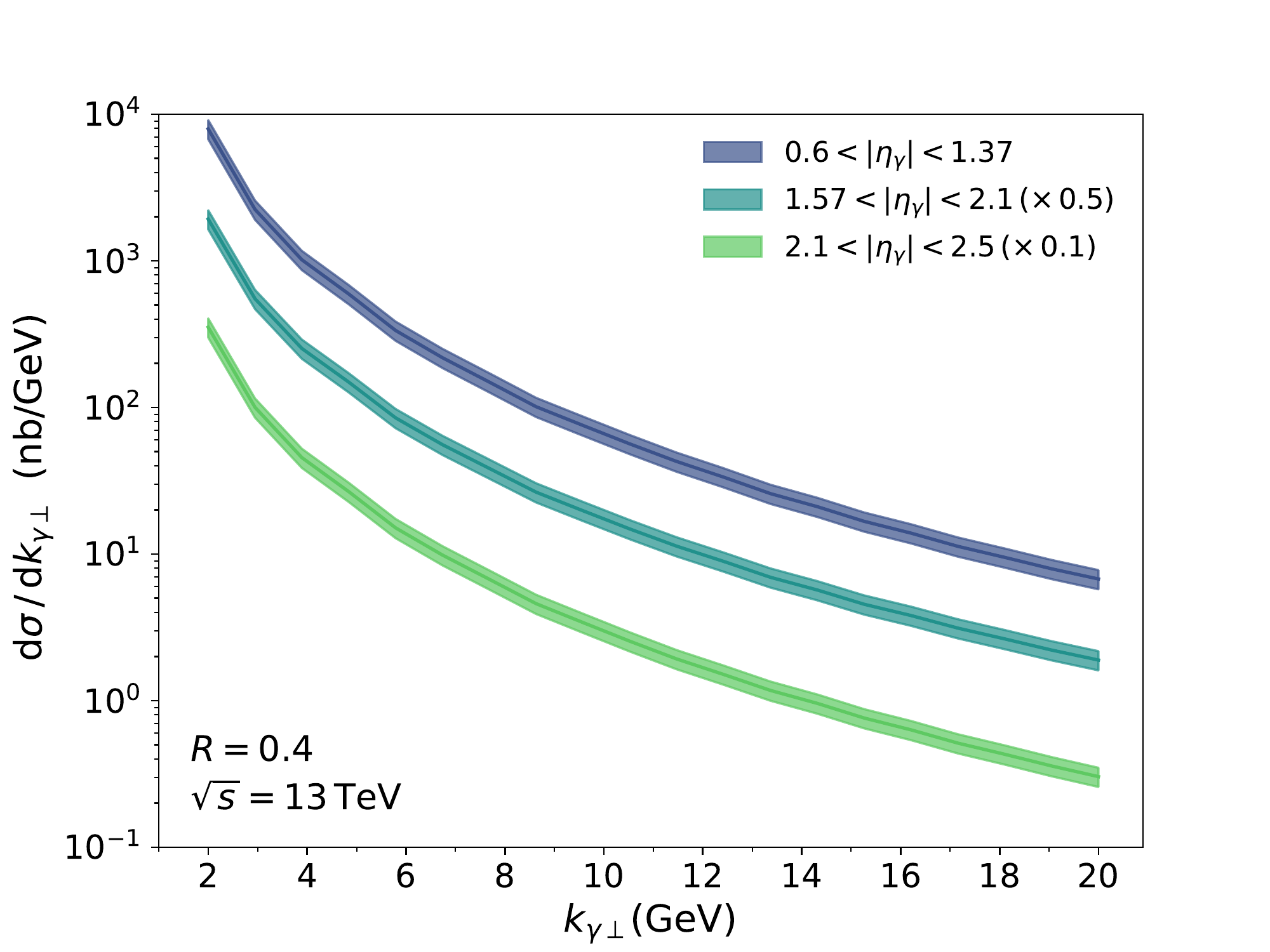}
  \end{center}
  \caption{Predictions for the inclusive photon production at $\sqrt{s}=13$ TeV across several rapidity bins. The central lines have the same $K$-factor as Fig.~\ref{fig:atlascms7TeV}.
  }
  \label{fig:13TeV}
\end{figure}

In Fig.~\ref{fig:atlascms7TeV} (Fig.~\ref{fig:13TeV}), we show the numerical results for the inclusive photon cross section based on Eqs.~\eqref{eq:inclo} and \eqref{eq:incnlo} at $7$ TeV ($13$ TeV) integrating over several $\eta_\gamma$ ranges up to $|\eta_{k_\gamma}| < 2.5$. In particular, we are covering the mid-rapidity region that can be measured by the LHC experiments.  The particular  rapidity 
ranges shown are those where ATLAS and CMS data exist presently at higher values of $k_{\gamma\perp}$. These data sets are for the CMS p+p data at 2.76 TeV \cite{Chatrchyan:2012vq} and at 7 TeV \cite{Chatrchyan:2011ue} for values $k_{\gamma\perp} \geq 20 \GeV$.  The ATLAS p+p data  set is given for 7 TeV, where one data point exists below $k_{\gamma\perp} = 20 \GeV$.
We have chosen the central value of this lowest lying ATLAS point in order to normalize our results and found that the required $K$-factor is $K = 2.4$. Interestingly, this is very close to the $K$-factor of 2.5 extracted in computations of $D$-meson production in this dilute-dense CGC framework~\cite{Ma:2018bax}. We have not shown a comparison to data above $k_{\gamma\perp}=20 \GeV$ because 
the contribution of logs in $k_\perp$ begin to dominate significantly over logs in $x$ around these values of $k_{\gamma\perp}$; the systematic treatment of these is beyond the scope of the present 
computation.

We have presented in this work an important first step towards constraining the proton UGDs at small-$x$ from inclusive photon production at the LHC. We can summarize our results as follows. We have quantified for the first time the dominant contributions to inclusive photon production at LO and NLO.  We found that the contribution of the NLO channel is significantly larger than the LO at central rapidities at the LHC. This is because at LHC energies the results are sensitive to small-$x$ values in the proton that have high gluon occupancy. We showed further that coherent rescattering contributions in the CGC that break $k_\perp$-factorization are at most about $10\%$ in the low $k_{\gamma\perp}$ region and negligible beyond $k_{\gamma\perp} \simeq 20$ GeV.
We have provided several numerical results for the inclusive isolated photon cross section that can be tested at the LHC.
Future publications will extend the analysis presented here to make predictions for p+A collisions and high multiplicity p+p and p+A collisions, and examine as well their sensitivity to available HERA dipole model fits~\cite{Albacete:2010sy}. Prior studies have only considered LO contributions to inclusive photon production. Another important avenue where progress is required is in the computation of higher order effects which formally are NNLO in this approach but are essential to quantify running coupling corrections and for matching to results from collinear factorization computations at high $k_{\gamma\perp}$ \cite{dEnterria:2012kvo,Campbell:2018wfu}.

{\bf Acknowledgements}
We thank Kevin Dusling for providing us his rcBK code. S.~B.\ thanks Yoshitaka Hatta for discussions and Davor Horvati\' c and Nenad Miji\' c for their help with the numerical procedures. S.~B. is grateful for the hospitality extended to him during his visit to BNL. S.~B.~is supported by a Japan Society for the Promotion of Science (JSPS) postdoctoral
fellowship for foreign researchers under Grant No. 17F17323. S.~B.~was previously supported by the European Union Seventh Framework Programme (FP7 2007-2013) under
grant agreement No. 291823, Marie Curie FP7-PEOPLE-2011-COFUND NEWFELPRO Grant No. 48. S. B. also acknowledges the support of the HRZZ Grant No. 8799 for computational resources. O. G. wants to thank J.~Berges  and A.~Mazeliauskas for discussions. K.~F.\ was supported by JSPS KAKENHI Grant No.\ 18H01211. R.~V's work is supported by the U.S. Department of Energy, Office of Science, Office of Nuclear Physics, under Contracts No. DE-SC0012704 and within the framework of  the TMD Theory Topical Collaboration. This work is part of and supported by the DFG Collaborative Research Centre ”SFB 1225 (ISOQUANT)”.




\bibliographystyle{utphys}
\bibliography{references}


\end{document}